\documentclass[12pt]{article}
\pdfoutput=1
\usepackage{jheppub}
\usepackage{amsmath,amsfonts,braket,amssymb,bm,bbm,xcolor,graphicx}
\usepackage[small,labelfont=bf]{caption}
\usepackage{array}
\usepackage{relsize}
\usepackage{mathtools}
\usepackage{braket}

\usepackage{enumerate}
\usepackage{caption}
\usepackage{subcaption}
\usepackage{comment}
\usepackage{tikz}

\allowdisplaybreaks

\usepackage[toc,page]{appendix}

\title{Cluster algebras for cosmological correlators}
\author[a]{Pouria Mazloumi,}
\author[b]{and Xiaofeng Xu}

\affiliation[a]{PRISMA$^+$ Cluster of Excellence \& Mainz Institute for Theoretical Physics\\
   Johannes Gutenberg University, Staudingerweg 7, 55099 Mainz, Germany}
\affiliation[b]{Department of Physics, Xiamen University, Xiamen, 361005, China}

\emailAdd{pmazloumi@uni-mainz.de}
\emailAdd{xuxiaofeng@xmu.edu.cn}

\abstract{In this paper,we explore the cluster algebras for symbol letters or singularities of cosmological correlators in a conformally coupled scalar field theory. We show that the symbol letters for tree-level $n$-site  ladder cosmological correlators are governed by $A_{2(n-1)}$ cluster algebras. Additionally, we demonstrate that the symbol letters for one-loop bubble cosmological correlator are an union of two $A_3$ cluster algebras. The algebras relations of letters will provide an important tool to bootstrap analytic cosmological correlators.  }

\begin{document} 
\begin{flushright}
{\small
MITP-25-081\\
}
\end{flushright}

\maketitle

\newpage

\section{Introduction}

In recent years, there has been enormous progress in understanding of mathematical structures on scattering amplitudes of quantum field theories (QFTs). Cluster algebras\cite{fomin2001clusteralgebrasifoundations,Fomin_2003,berenstein2004clusteralgebrasiiiupper}, introduced by Fomin and Zelevinsky, play an important role in the study of scattering amplitudes both at the level of integrand and function space of analytical expression, especially in $N = 4$ super-symmetric Yang-Mills (SYM) theory. 

It was first discovered in \cite{Arkani-Hamed:2012zlh} that the all-loop planar integrands in $\mathcal{N} = 4$ SYM are encoded by positive Grassmannian $Gr^{\geq 0}(k,n)$. This geometric object, intimately linked to Grassmannian cluster algebras of type $Gr(k,n)$, provides a combinatorial framework for organizing scattering amplitudes. Remarkably, subsequent research in  \cite{Golden_2014,Golden:2014xqa} revealed that the integrated scattering amplitudes are expressed in terms of cluster polylogarithm functions, whose symbol letters are the $\mathcal{A}$ coordinates of Grassmannian. 
These functions exhibit a very special property, known as "cluster adjacency"\cite{Drummond:2017ssj}, which is closely related to the extended Steinmann relations\cite{Caron-Huot:2016owq,Dixon:2016nkn, Caron-Huot:2019bsq, Caron-Huot:2020bkp}. It indicates how different singularities are related to each other, and puts strong constraints on scattering amplitudes. Due to the constraints, it makes cluster algebras an important tool for  bootstrapping amplitudes to very high loop orders \cite{Dixon:2011pw,Dixon:2014xca,Dixon:2014iba,Drummond:2014ffa,Dixon:2015iva,Caron-Huot:2016owq,Dixon:2016nkn,Drummond:2018caf,Caron-Huot:2019vjl,Caron-Huot:2019bsq,Dixon:2020cnr,Caron-Huot:2020bkp,Dixon:2020bbt,Guo:2021bym,Dixon:2022rse,Papathanasiou:2022lan,He:2025tyv}.

Given the established relevance of cluster algebras to $\mathcal{N}=4$ SYM through Grassmannian cluster structures, it is natural to inquire whether they also feature in broader QFTs. Recently, an increasing body of evidence suggests cluster algebras are closely related to the singularities of Feynman integrals in general QFTs. In \cite{Chicherin:2020umh}, the authors show that the two-loop four-point Feynman integrals with one off-shell leg are described by a $C_2$ cluster algebra. The corresponding Feynman integrals are associated to next-to-next-to-leading-order (NNLO) quantum chromodynamics (QCD) corrections for the processes of vector boson or Higgs plus jet production. Furthermore, it's discovered in \cite{Aliaj:2024zgp} that the symbol letters of non-planar three loop Feynman integrals in these processes are governed by $G_2$ cluster algebra, which is an extension of $C_2$ cluster algebra. Further examples of cluster algebras for Feynman integrals, including their relations to Schubert problems and Grassmannian cluster algebras, are explored in  \cite{He:2021esx,He:2021fwf,He:2021non,He:2021mme,He:2021zuv,He:2021eec,He:2022ctv,He:2022tph,Zhao:2023okw,Pokraka:2025ali,He:2024fij}.

In this paper, we would like to extend the study of cluster algebras to cosmological correlators\cite{Arkani-Hamed:2015bza,Arkani-Hamed:2023bsv,Arkani-Hamed:2023kig,Baumann:2024mvm, Hang:2024xas,De:2023xue,Fevola:2024nzj,Qin:2024gtr,De:2024zic,Capuano:2025ehm,Glew:2025ypb,Baumann:2025qjx}. We will investigate the cluster algebras appearing in the singularities or symbol letters of  cosmological correlators in a conformally coupled scalar filed theory. In \cite{Arkani-Hamed:2023bsv,Arkani-Hamed:2023kig,Baumann:2024mvm}, the authors develop a formalism to derive differential equations for the basis functions of these correlators, which are represented by tubing graphs. Using operations—including activation, grow, merge, and absorption—one can directly write down these differential equations. Within this formalism, singularities are also represented as tubing graphs. By examining the form of these singularities or symbol letters, we will show that they are closely related to the $A_n$ cluster algebras. 

The paper is organized as follows: in Section 2, we review the basic concepts of cluster algebras and cluster polylogarithms, focusing on $A_n$ cluster algebras and their associated cluster letters; in Section 3, we explore the symbol letters of tree level ladder cosmological correlators in a conformally coupled scalar field theory and show that the letters are  described by $A_n$ cluster algebras; in section 4, we study the cluster algebras for one-loop bubble cosmological correlators; in section 5, we summerize our results and conclude.

\textbf{Note.} During the final preparation of this manuscript we made aware of another paper \cite{Livia} with the similar content. We would like to thank the authors of that paper for the coordination of the publications. 

\section{Review of Cluster algebras}

\subsection{Cluster algebras and cluster polylogarithms}

Cluster algebras are subrings of rational functions generated by combinatorial data. The main object in the cluster algebras is the notion of seeds, which consist of a set of algebraically independent generators $x_i$, or the cluster variables, and an exchange matrix $B$. The generators are grouped into a subset of rank $n$, $\mathbf{x} \equiv \{x_1, \dots, x_n\}$, known as the cluster of the seed or the $\mathcal{A}$-coordinates. The seed is denoted as a pair $(\mathbf{x}, B)$. Starting with the initial seed and through processes called mutations, the seed $(\mathbf{x}, B)$ evolves by replacing a cluster variable and altering the exchange matrix, generating a new seed $( \mathbf{x}^\prime, B^\prime) $. Repeating this mutation process generates the entire cluster algebra. A cluster may also include frozen variables or coefficients $\{x_{n+1}, \dots, x_m\}$, which is invariant under the mutations. 

The mutation is encoded by the exchange matrix $B$.  It's a  $m\times n$ integer matrix and  the principal part of the exchange matrix $\tilde{B}=(b_{ij})_{1\le i, j \le n}$  is a skew-symmetric matrix, where  $b_{ij}$ are the element components of matrix $B$. The mutation at k-th variables, $\mu_k(\mathbf{x}, B)= (\mathbf{x}^\prime, B^\prime), 1\le k \le n, $ is defined as follows: the cluster $\mathbf{x}^\prime$ is equal to $\mathbf{x}$ except for $x_k$  is replaced by $x^\prime_k$, satisfying the relation
\begin{equation}
   \mathbf{x}'=     \begin{cases}
    &  x^\prime_j=x_j,  \qquad \qquad j\neq k  \\
    &  x^\prime_j =\frac{1}{x_j} \prod_{i=1}^m x_i^{[b_{ij}]_+}+\prod_{i=1}^m x_i^{[-b_{ij}]_+}, \quad \quad i=k
    \end{cases}
\end{equation}
where $[a]_+=\max(0,x)$.
The exchange matrix $ B^\prime$ is generated via the following: 
\begin{equation}
  b^\prime_{ij} =
    \begin{cases}
      -b_{ij} & \text{if $i=k$ or $j=k$}\\
      b_{ij} +[b_{ik}]_+ b_{k j} +[-b_{ik}]_+ b_{k j} & \text{otherwise} \, .
    \end{cases}       
\end{equation}

\begin{figure}
\centering
\begin{tabular}{cc}
$\includegraphics[scale = 0.5, keepaspectratio = 0.3]{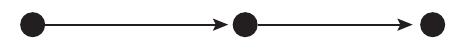} $ & \quad 
$\begin{pmatrix}
0 & 1 & 0  \\
-1 & 0 & 1  \\
0 & -1 & 0   \nonumber
\end{pmatrix} $ \\
$ $ $\quad \ \ \downarrow \text{$\mu_2$} $ \\ 
$\includegraphics[scale = 0.5, keepaspectratio = 0.3]{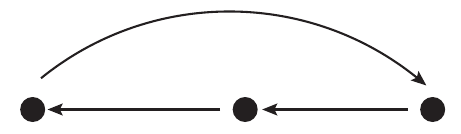} $ & \quad 
$\begin{pmatrix}
0 & -1 & 1  \\
1 & 0 & -1  \\
-1 & 1 & 0   \nonumber
\end{pmatrix} $ \\
$ $ $\quad \ \ \downarrow \text{$\mu_3$} $ \\ 
$\includegraphics[scale = 0.5, keepaspectratio = 0.3]{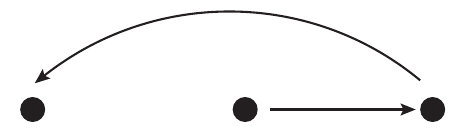} $ & \quad 
$\begin{pmatrix}
0 & 0 & -1  \\
0 & 0 & 1  \\
1 & -1 & 0   \nonumber
\end{pmatrix} $ 
\end{tabular}
\caption{An example of a quiver mutation. }
\label{fig:quiver_example}
\end{figure}

It's often quite convenient to use a quiver or an oriented diagrams (without 2-cycles, i.e. arrow like $a \rightarrow b \rightarrow a $ ), to represent the exchange matrix. Let the vertices correspond to the rows and columns in the exchange matrix, then the absolute value of element  $|b_{ij}|$ in exchange matrix corresponds to the number of arrows connecting $i$ and $j$. If $b_{ij}>0$ the arrow is from $i$ to $j$; otherwise, arrows are from $j$ to $i$.  The mutation of k-th variable corresponds to the mutation of the quiver at a vertex k which is done in the following steps: First, for all paths of the form $i \rightarrow k \rightarrow j$, add an arrow from $i$ to $j$ and remove the 2-cycles. Second, reverse all arrows incident with $k$. An example of the mutation of a quiver is shown in fig.~(\ref{fig:quiver_example}).

In general, the mutation of quiver will generate infinite number of seeds. However there exist certain types of cluster algebras such that they have finitely many seeds. It was proven in \cite{Fomin_2003} that the finite type of cluster algebras are those whose quiver (specifically the mutable part of the quiver) can be mutated from a Dynkin diagram of type A,B, C, D, E, F,G. The number of cluster variables N for these cluster algebras are:
\begin{align}
& N(A_n) = \frac{n(n+3)}{2} \, , \quad N(B_n) =N(C_n)= n(n+1) \, , \quad N(D_n)=n^2 \, , \nonumber \\
& N(E_6)=42, \, , \quad N(E_7)=70 \, , \quad N(E_8)=128 \, , \quad  N(F_4)=28 \, , \quad  N(G_2)=8  \, .
\end{align}

Furthermore, as shown in \cite{Golden_2014,Golden:2014xqa}, there is a natural polylogarithm function space associated to a finite type of cluster algebra, which is  determined by the corresponding $\mathcal{A}$-coordinates. The weight-$w$ cluster polylogarithm functions are defined by having the following properties
\begin{align}
d I^{(w)} = \sum_ i I_i ^{(w-1)} d \log x_i \, ,
\label{eq: cluster_dif}
\end{align}
where $I_i ^{(w-1)}$ is a weight-$(w-1)$ cluster polylogarithm function and $x_i$ are the coordinates of the cluster algebras. These weight-$w$ functions can be expressed as Chen iterated integrals. 

Similar to the Eq.~(\ref{eq: cluster_dif}), the differential equations of cosmological correlators in a conformally coupled scalar field theory satisfies following $\epsilon$-factorized form \cite{Arkani-Hamed:2023bsv,Arkani-Hamed:2023kig,Baumann:2024mvm, De:2023xue}
\begin{align}
d \vec{f} (\bm{z}, \epsilon) = \epsilon \sum_ i \bm{A}_i  d \log x_i(\bm{z})  \vec{f} (\bm{z}, \epsilon) \, ,
\end{align}
where $\bm{z}$ are kinematic variables and $\epsilon$ is a parameter associated to different space times. The kernel of $d\log$ or $x_i(\bm{z})$ is called a letter, and the set of letters is called alphabet. The solutions admit a series expansion in $\epsilon$, with coefficients given by Chen iterated integrals. Notably, the letters $x_i(\bm{z})$ play a similar role to $\mathcal{A}$-coordinates in cluster algebras and determine the function space of solutions. Motivated by this correspondence, we would like to investigate cluster algebras for the letters arising in these differential equations.

\subsection{$A_n$ cluster algebras}
\label{sec:A_n cluster algebras}

As we will show, the letters of differential equations of cosmological correlators in a conformally coupled scalar field theory are closely related to $A_n$ cluster algebras. Therefore we discuss the combinatorial model for the $A_n$ cluster algebras in detail, which is introduced in \cite{Fomin_2003}.  

\begin{figure}[h]
    \centering
    \includegraphics[scale=0.8]{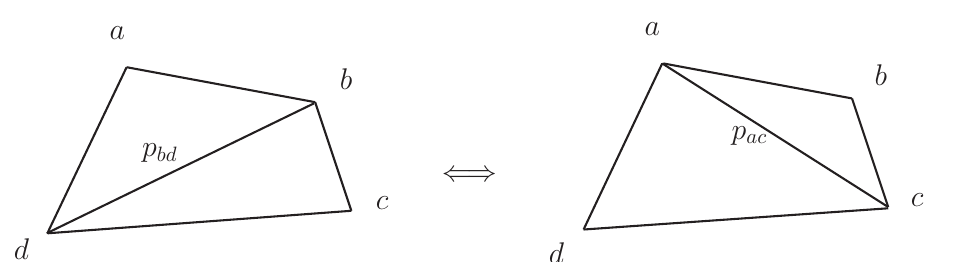}
    \caption{Triangulation flip of a diagonal in quadrilateral. }
    \label{fig:triangular-flip}
\end{figure}

Let $\mathbf{P}_N$ be a regular N-gon with vertices labeled in the counterclockwise direction, and $p_{ij}$ denotes length of an edge or a diagonal between vertex $i$ and $j$. A  \textit{triangulation} of $\mathbf{P}_N$ is a maximal set of non-crossing diagonals of $\mathbf{P}_N$, which divide the $N$-gon into triangles. Different triangulations can be related by the so called \textit{triangulation flip} operation, which replaces one of the diagonal in quadrilateral by the other, as shown in fig.~\ref{fig:triangular-flip}. The diagonals in quadrilateral satisfies \textbf{Ptolemy’s Theorem}, it states that for a cyclic quadrilateral, the product of the diagonals equals the sum of the products of the two pairs of opposite sides.
\begin{align}
p_{bd} p_{ac}=p_{ab} p_{cd} +p_{bc} p_{ad} \, .
\label{eq:ptolemy}
\end{align} 

It is proven in \cite{Fomin_2003} that for an $A_n$ cluster algebra with $\mathcal{A}$-coordinates, there exists a bijection between the cluster variables and the diagonals of a triangulation of the polygon $\mathbf{P}_{n+3}$, while the frozen variables correspond to the boundary edges of $\mathbf{P}_{n+3}$. Mutations correspond to \textit{flipping} diagonals in the triangulation. This relationship can be interpreted through an associated quiver: the diagonals in the triangulation correspond to mutable vertices in the quiver, the boundary edges correspond to frozen vertices, and the arrows in the quiver are determined by the clockwise orientation of the triangle's boundaries.

\begin{figure}[h]
    \centering
    \includegraphics[scale=0.5]{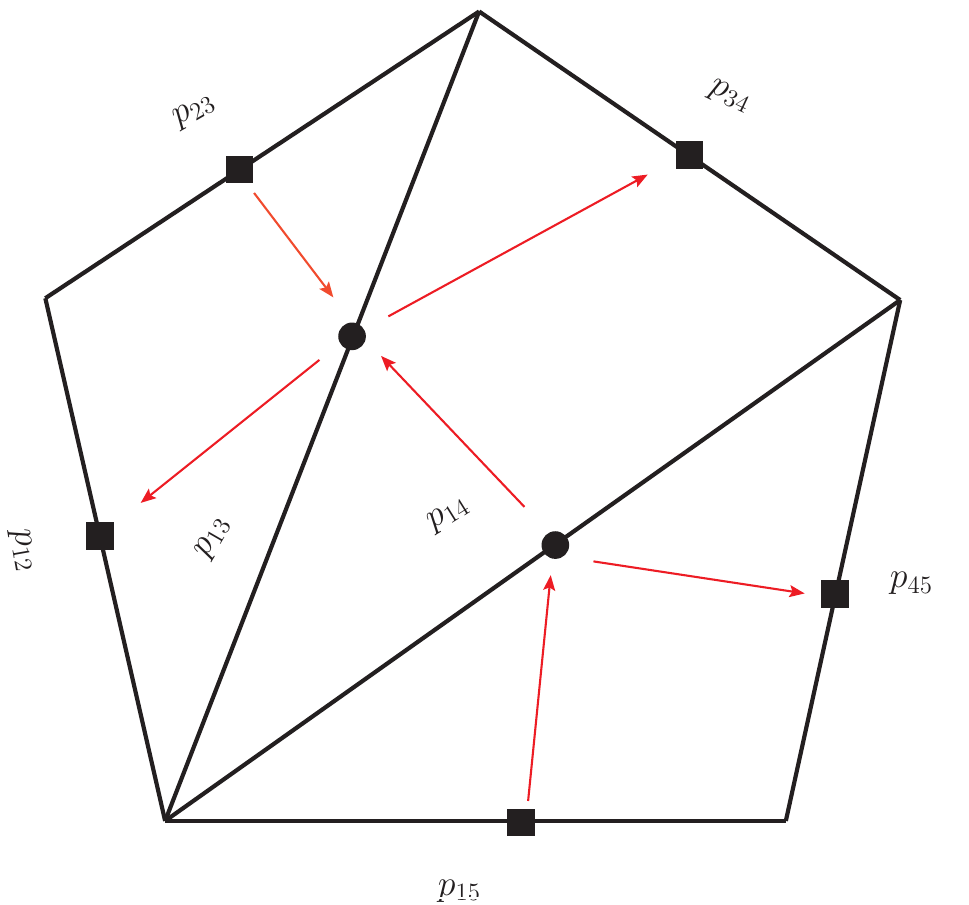}
    \caption{Realization model of $A_2$ cluster algebras with a 5-gon, where the dots represent mutable vertices, while the squares denote frozen variables  }
    \label{fig:example-5-gon}
\end{figure}

As an example we take a look at $A_2$ cluster algebras. It is related to the triangulation of 5-gon. We can  choose $p_{13}$ and $p_{14}$ as the diagonal of triangulation, and then the edges $p_{12}, p_{23},p_{34},p_{45}$ and $p_{1,5}$ are frozen variables. The quiver is shown in fig.~\ref{fig:example-5-gon}. 

In this paper, we focus on the cluster algebra of letters appearing in the differential equations of cosmological correlators, in which the letters are closely related to the $A_n$ cluster algebras. For later convenience, here we list the cluster letters for an $A_n$ cluster algebra 
\begin{align}
\Phi_n= \bigcup_{i=1}^{n} \{ z_i, 1+z_i \} \cup \bigcup_{1\le i\le j\le n } \{ z_i-z_j \} \, .
\end{align}
which can be obtained by rational transformation of the corresponding $\mathcal{A}$-coordinates \cite{Arkani-Hamed:2020tuz}.

\section{Cluster algebras for tree-level ladder cosmological correlators}

\subsection{Cluster algebras for the correlators}

\begin{figure}[h]
    \centering
    \includegraphics[scale=0.8]{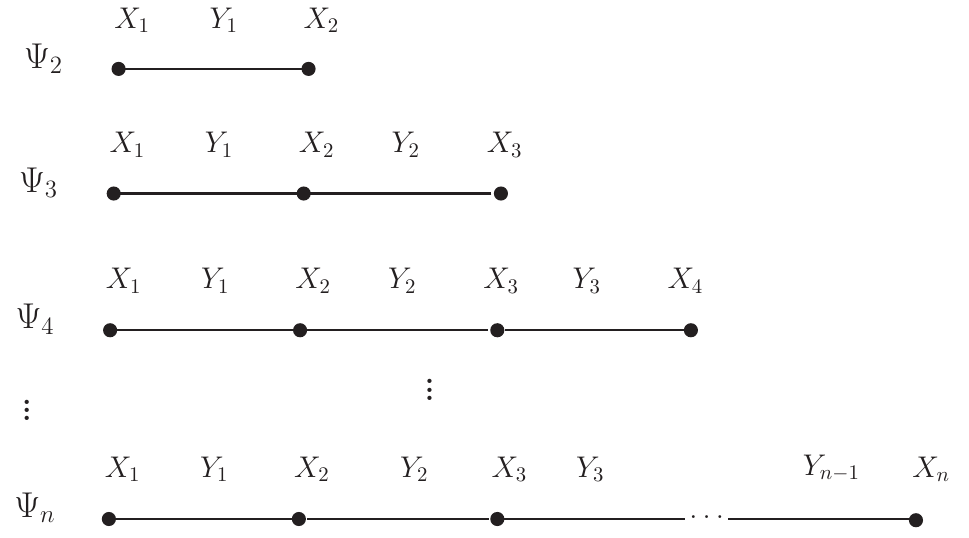}
    \caption{Tree-level n-site ladder cosmological correlators.}
    \label{fig:ladder-correlators}
\end{figure}

In this subsection, we will discuss the cluster algebra structure of letters appearing in the differential equations of tree-level ladder cosmological correlators. The corresponding diagrams are depicted in fig.~(\ref{fig:ladder-correlators}). Following the rules of kinematic flow in \cite{Arkani-Hamed:2023bsv,Arkani-Hamed:2023kig}, one can directly read the symbol letters of cosmological correlators. 
 
We first consider the alphabet for the two-site correlators, as shown in the first diagram in fig.~\ref{fig:ladder-correlators}. It has 5 letters given by,  
\begin{align}
\Phi_2= \left\{X_1-Y_1,X_1+Y_1, X_2-Y_1, X_2+Y_1, X_1+X_2\right\} \, . 
\label{eq: phi_2}
\end{align}
In order to identify the cluster algebra structure of the letters, we perform the following transformation
\begin{align}
z_1=\frac{2Y_1}{X_1-Y_1} \, , \quad z_2=\frac{X_2+Y_1}{X_1-Y_1} \, .
\end{align} 
The new alphabet is given by
\begin{align}
\tilde{\Phi}_2= \left\{z_1,1+z_1,z_2,1+z_2,z_1-z_2\right\} \, . 
\end{align}
The transformation introduces $2Y_1$ as an additional dimensional letter. However, since the final correlators are dimensionless functions (modulo an overall dimensional factor), only the dimensionless letters in $\tilde{\Phi}_2$ are physically relevant. One can readily check that the alphabet $\tilde{\Phi}_2$ has the exact form of $A_2$ cluster algebra.

Next we consider three-site correlators. There are 13 letters in the differential equations: 
\begin{align}
  \Phi_3=  \{ &X_1-Y_1,X_1+Y_1,  X_3-Y_2, X_3+Y_2, X_2+X_3-Y_1,X_2+X_3+Y_1, \,  \nonumber
  \\
   &X_1+X_2-Y_2, X_1+X_2+Y_2,  X_2-Y_1-Y_2, X_2+Y_1-Y_2,X_2-Y_1+Y_2, \, \nonumber
   \\
   &X_2+Y_1+Y_2, X_1+X_2+X_3 \} \, .
\end{align}
Similar to two-site correlators, we define new variables as
\begin{align}
z_1& =\frac{2Y_1}{X_1-Y_1} \, , \quad  z_2=\frac{X_2+Y_1-Y_2}{X_1-Y_1} \, , \quad  z_3=\frac{X_2+Y_1+Y_2}{X_1-Y_1}  \, , \nonumber
\\
z_4&=\frac{X_2+X_3+Y_1}{X_1-Y_1} \, .
\end{align}
The alphabet will transformed to 
\begin{align}
 \tilde{ \Phi}_3=  \left\{z_1,1+z_1,z_2,1+z_2,z_3,1+z_3,z_4,1+z_4,z_1-z_2,z_1-z_3,z_1-z_4,z_2-z_4,z_3-z_4\right\} \, ,
\end{align}
which is the $A_4$ cluster algebra except a missing letter $z_2-z_3$. 

We can perform similar transformation for the alphabet of four-site correlators, which is given by:
\begin{align}
\Phi_4=&\left\{X_1-Y_1,X_1+Y_1, X_4-Y_3,X_4+Y_3, X_1+X_2+X_3+X_4,  \right. \, \nonumber
\\
&\left. X_1+X_2-Y_2,X_1+X_2+Y_2,X_3+X_4-Y_2,X_3+X_4+Y_2,   \right. \, \nonumber
\\
& \left. X_2-Y_1-Y_2,X_2+Y_1-Y_2, X_2-Y_1+Y_2, X_2+Y_1+Y_2,\right. \, \nonumber
\\
&\left. X_3-Y_2-Y_3,X_3+Y_2-Y_3, X_3-Y_2+Y_3, X_3+Y_2+Y_3, \right. \, \nonumber 
\\
& \left. X_2+X_3-Y_1-Y_3,X_2+X_3+Y_1-Y_3, X_2+X_3-Y_1+Y_3,X_2+X_3+Y_1+Y_3,\right. \, \nonumber
\\
& \left. X_1+X_2+X_3-Y_3,  X_1+X_2+X_3+Y_3 ,X_2+X_3+X_4-Y_1, X_2+X_3+X_4+Y_1 \right\} \, .
\end{align}
We obtain the following new alphabet for four-site correlators
\begin{align} 
\tilde{\Phi}_4=&\left\{  z_1,1+z_1,z_2,1+z_2,z_3,1+z_3,z_4,1+z_4,z_5,1+z_5,z_6,1+z_6,z_1-z_2,z_1-z_3, \right.  \, \nonumber
\\
&\left.  z_1-z_4,z_1-z_5,z_1-z_6,z_2-z_4,z_2-z_5,z_2-z_6,z_3-z_4,z_3-z_5,z_3-z_6, \right.  \, \nonumber
\\
& \left. z_4-z_5,z_4-z_6,z_5-z_6\right \} \, ,
\end{align}
where the variables are defined as
\begin{align}
z_1& =\frac{2Y_1}{X_1-Y_1} \, , \quad  z_2=\frac{X_2+Y_1-Y_2}{X_1-Y_1} \, , \quad  z_3=\frac{X_2+X_3+Y_1-Y_3}{X_1-Y_1}  \, , \nonumber
\\
z_4&=\frac{X_2+X_3+Y_1-Y_3}{X_1-Y_1} \, , \quad z_5=\frac{X_2+X_3+Y_1+Y_3}{X_1-Y_1} \, , \quad 
z_6=\frac{X_2+X_3+X_4+Y_1}{X_1-Y_1}  \, .
\end{align}
We can see that the alphabet is almost the $A_6$ cluster algebra, except for the missing letters $z_2-z_3$ and $z_4-z_5$. 

Similarly, we find that the alphabet of five-site correlator after variable transformation is almost the $A_8$ cluster algebra modulo the missing letters $z_2-z_3$,  $z_4-z_5$ and  $z_5-z_6$. These examples reveal a consistent pattern: $n$-site correlator letters form $A_{2(n-1)}$-type cluster algebras with specific letter exclusions.

\subsection{Geometry of the letters}

As discussed in Section.~(\ref{sec:A_n cluster algebras}),  there exists a bijection between cluster algebras of type $ A_n $ and triangulations of $n+3$-gon $\mathbf{P}_{n+3} $. Since the letters of n-site correlators correspond to $A_{2(n-1)} $-type cluster algebras, we aim to investigate the letters appearing in the cosmological correlators geometrically with regular (2n+1)-gon.  

\begin{figure}[h]
    \centering
    \includegraphics[scale=1]{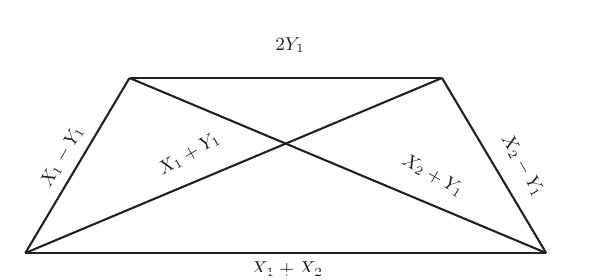}
    \caption{Quadrilateral associated to the letters of two-sites correlators.}
    \label{fig:quadrilateral}
\end{figure}

\begin{figure}[h]
    \centering
    \includegraphics[scale=0.6]{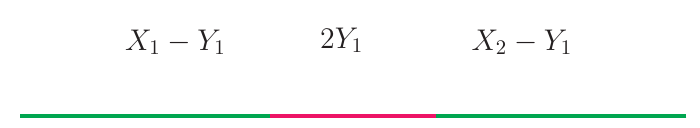}
    \caption{The quadrilateral can be viewed as a partion of a line segment of length $(X_1+X_2)$.}
    \label{fig:2-segment}
\end{figure}

Let's first consider the associated letters appearing in the two-site correlators as shown in eq.~(\ref{eq: phi_2}). 
Among these letters, we observe following identity:  
\begin{align}
(X_1-Y_1)(X_2-Y_1)+2Y_1(X_1+X_2)=(X_1+Y_1)(X_2+Y_1) \, .
\end{align}
This relation resembles the Ptolemy's relations in eq.~(\ref{eq:ptolemy}), which states that for a cyclic quadrilateral, the product of the diagonals equals the sum of the products of the two pairs of opposite sides. Therefore, we can construct the following quadrilateral and assign the letters to the edges and diagonals, as shown in fig.~(\ref{fig:quadrilateral}). However, noticing that the sum of it's edges $(X_1-Y_1)$, $2Y_1$, and $(X_2-Y_1)$ equals $(X_1+X_2)$, the quadrilateral can be interpreted as a partition of a line segment of length $(X_1+X_2)$, or the total energy of external legs, as shown in fig.~(\ref{fig:2-segment}).

In addition, we have following relations among the letters 
\begin{align}
&(X_1-Y_1)+(X_2+Y_1)=(X_1+X_2) \, \nonumber \, , 
\\
&(X_1+Y_1)+(X_2-Y_1)=(X_1+X_2) \, ,
\label{eq: additional-relation}
\end{align}
which is not captured by the quadrilateral above. In order to include these relations, we embed the quadrilateral to a 5-gon, shown in fig.~(\ref{fig:2-segment}).
\begin{figure}[h]
    \centering
    \includegraphics[scale=0.5]{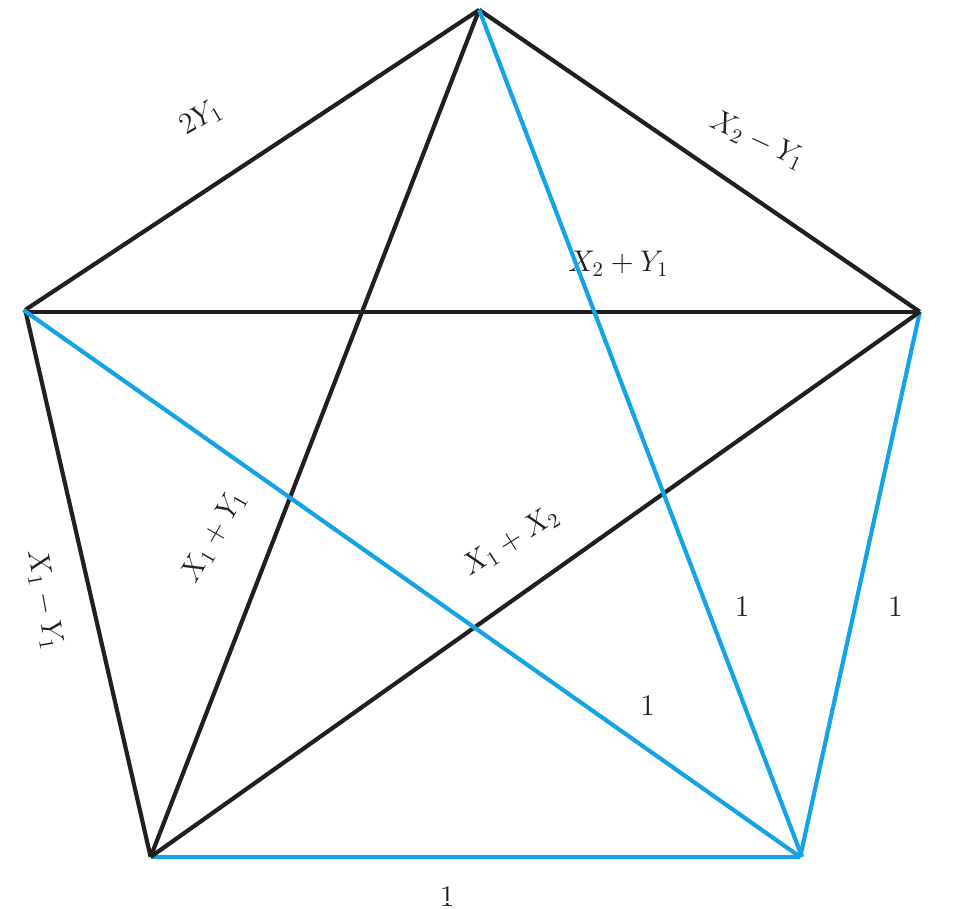}
    \caption{Embedding the letters of two-sites correlators into a 5-gon.}
    \label{fig:5-gon}
\end{figure}
The length of the edges or diagonals associated to the new vertex is assigned to 1. We can see that quadrilaterals consisting the new vertex inside the 5-gon are compatible with the relations in eq.~(\ref{eq: additional-relation}). 

\begin{figure}[h]
    \centering
    \includegraphics[scale=0.5]{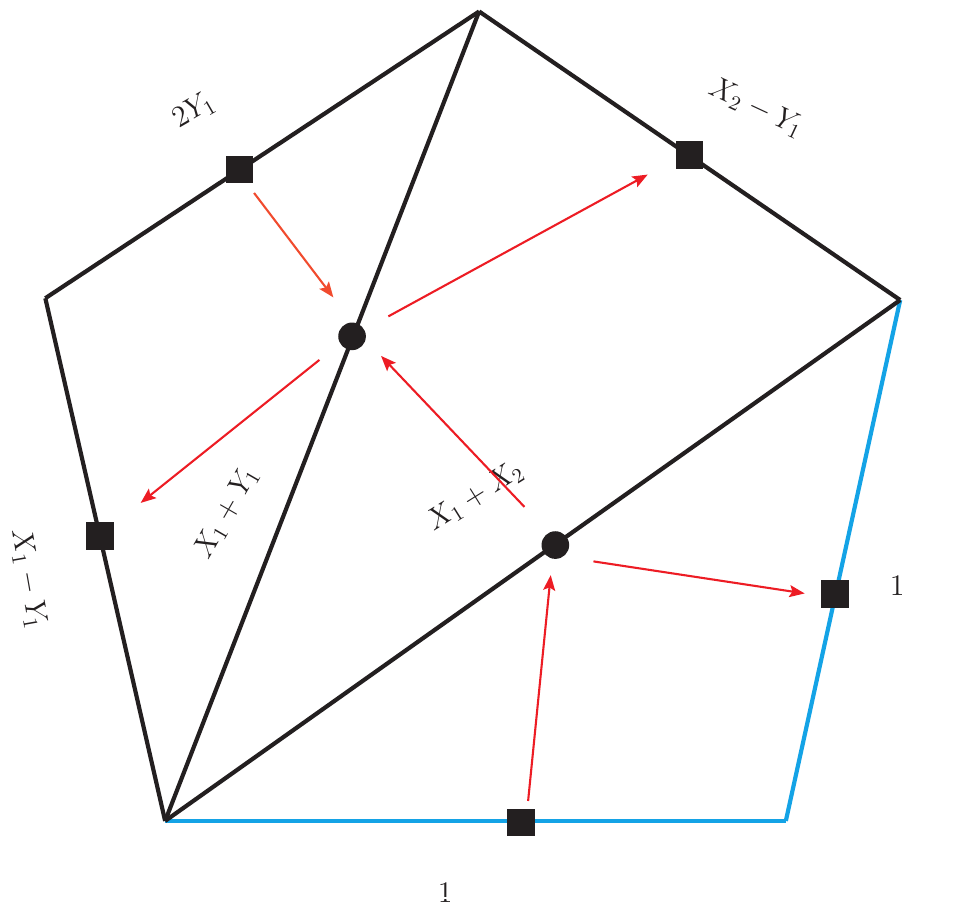}
    \caption{Quiver for the letters of two-sites correlators.}
    \label{fig:tri-5-gon}
\end{figure}

Due to the fact that there is a correspondence of $A_n$ type of cluster algebras and triangularization of (n+3)-gon, we can choose following triangularization as initial seed, as shown in  fig.~(\ref{fig:tri-5-gon}). 
The cluster $\mathcal{A}$-coordinates are $\{X_1+Y_1,X_1+X_2 , X_1-Y_1, 2Y_1,X_2-Y_1,1,1\}$, where the last five entries are frozen coordinates or coefficients.  The arrows, determined by the clock orientation of the boundary of the triangles,  represent the quiver.

\begin{figure}[h]
    \centering
     \includegraphics[scale=0.6]{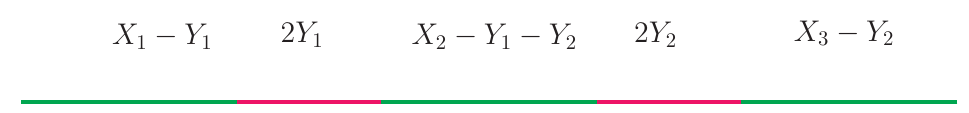} 
    \includegraphics[scale=0.5]{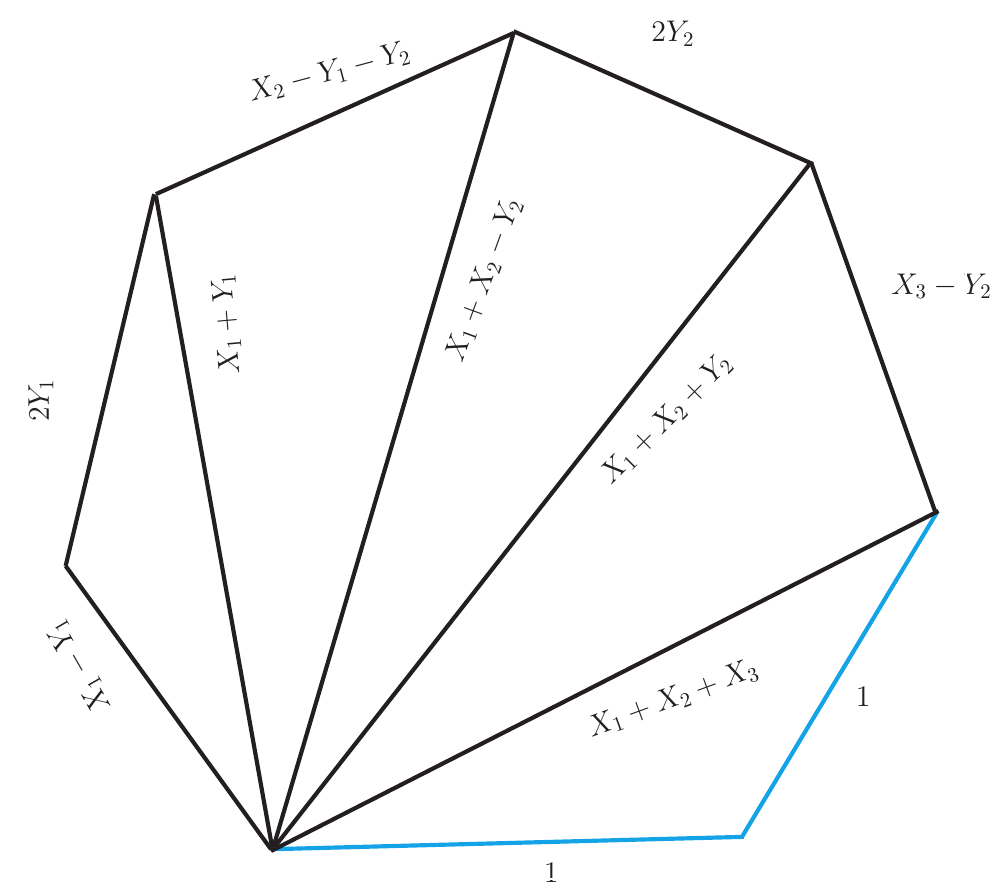}
    \caption{ Embedding the letters of three-sites correlators into a 7-gon.}
    \label{fig:7-gon}
\end{figure}

Following the same method, we can construct a 7-gon for the three-sites cosmological correlators, as depicted in fig.~(\ref{fig:7-gon}). The 6-gon, consisting the diagonal $(X_1+X_2+X_3)$ and all the other black edges, can be viewed as a partition of a line segment of length $(X_1+X_2+X_3)$. More generally, the letters for tree-level $n$-site correlators can be embedded to a $(2n + 1)$-gon  using the same approach. Therefore, we can see that these letters correspond to cluster algebras of type $A_{2(n-1)}$.  

So far, our discussion has focused on tree level ladder cosmological correlators. Therefore, it would be very interesting to explore the algebra relations of letters in correlators beyond tree level. In the Appendix~\ref{appendix:A}, we show that the letters of one-loop bubble cosmological correlators are related to cluster algebras of $A_3$, but leaving the discussion of generic case to future work.

\section{Conclusions and outlook}

In this paper, we initiate the study of cluster algebras in the context of cosmological correlators. We extract the symbol letters of these correlators by the method of kinematic flow. By examining the algebraic relations among the symbol letters, we find that the letters of $n$-sites tree level ladder cosmological correlators can be embedded into a $(2n+1)-$gon. This embedding reveals that the symbol letters exhibit the structure of a cluster algebra of type $A_{2(n-1)}$. Furthermore, the $(2n+1)-$gon admits a natural interpretation as a partition of a line segment, whose length corresponds to the total energy of the external legs. 

We also explore the case of one-loop bubble cosmological correlators. The symbol letters in this case consists of two subsets, each of which forms an $A_3$ cluster algebra, which is similar to the tree level ladder cosmological correlators. 

While our work focuses on ladder type cosmological correlators, it would be intriguing to investigate the cluster structures of star-like cosmological correlators, particularly whether their symbol letters admit a unified cluster algebra description or can be understood as a union of smaller cluster algebras, as in the one-loop bubble case. Furthermore, recently a block decomposition of the correlators using zonotops has been established in \cite{Glew:2025ypb,Baumann:2025qjx}. This decomposition may be used in our method to decompose the cluster algebra of one loop bubble or star graphs. Finally, extending this analysis to massive cosmological correlators offers another important direction for future exploration\cite{Gasparotto:2024bku,Chen:2024glu,Liu:2024str}.

In the future, we will further investigate the origin of cluster algebras in cosmological correlators, and explore potential connections with positive geometry. Additionally, it would be interesting to study the cluster adjacency in context to the $A_n$ type cluster algebras, which govern the relation of letters. This, in turn, can be used for bootstrapping analytic expressions for cosmological correlators.

\subsection*{Acknowledgements}

We are indebted to Stefan Weinzierl and Federico Gasparotto for several pieces of advice on different aspects of this work; we would like to thank Antonela Matijašić and Xuhang Jiang for useful discussions concerning cluster algebra. We are grateful to Andrzej Pokraka for comments and feedback on an earlier version of this draft.\\
This work has been supported by the Cluster of Excellence Precision Physics,
Fundamental Interactions, and Structure of Matter (PRISMA EXC 2118/1) funded by the German Research Foundation (DFG) within the German Excellence Strategy (Project ID 390831469). Xiaofeng Xu expresses his gratitude to Matthias Neubert. His postdoctoral research in Mainz is supported by funding from the European Research Council (ERC) under the European Union’s Horizon 2022 Research and Innovation Program (Grant Agreement No. 101097780, EFT4jets). 

\newpage

\appendix

\section{Cluster algebras for one-loop bubble cosmological correlators} 
\label{appendix:A}

In this section, we will investigate the algebra relations of letters appearing in the one-loop cosmological correlators. For the one-loop bubble cosmological correlator, it's written as
\begin{align}
\psi_{(2), \text{bubble}}= -\int_0^\infty  (\omega_1\omega_2)^ \epsilon  \frac{4Y_1 Y_2}{B_1 B_2 B_3} \left( \frac{1}{B_4}+\frac{1}{B_5} \right) d^2 \omega \, , 
\end{align}
where the factor $(4 Y_1 Y_2) $ is introduced for convenient. $B_i$ are hyperplanes defined as
\begin{align}
&B_1=\omega_1+X_1+Y_1+Y_2 \, \quad B_2=\omega_2+X_2+Y_1+Y_2 \, , \nonumber
\\
&B_3=\omega_1+\omega_2+X_1+X_2 \, \quad B_4=\omega_1+\omega_2+X_1+X_2+2Y_2 \, , \nonumber
\\
&B_5=\omega_1+\omega_2+X_1+X_2+2Y_1 \, . 
\end{align}
With the kinematic flow method\cite{Baumann:2024mvm}, we can read the alphabet
\begin{align}
\Phi_{(2), \text{bubble}}= &\left\{  X_1-Y_1-Y_2, X_1+Y_1-Y_2,X_1-Y_1+Y_2,X_1+Y_1+Y_2 \, , \right. \nonumber
\\
& \left. X_2-Y_1-Y_2, X_2+Y_1-Y_2,X_2-Y_1+Y_2,X_2+Y_1+Y_2 \right. \, ,\nonumber
\\
& \left. X_1+X_2, X_1+X_2+2Y_1,X_1+X_2+2Y_2 \right\} \, .
\end{align}
However we find that the algebraic relations among letters can not be understood in the same way as the tree-level ladder correlators, which is based on the partition of total energy of external legs.  We notice that the letter $X_1+X_2+2Y_1$ will never talk to $X_1+X_2+2Y_2$, that is to say these two letters do not appear in cluster functions simultaneously. It's because $X_1+X_2+2Y_2$ is associated to the hyperplane $B_4$ and $X_1+X_2+2Y_1$ to the hyperplane $B_5$, which correspond to two different parts of integrand in the correlators. Thus we may consider algebraic structures of subsets of the alphabet. 

By investigating the singularities appearing in the integral associated with hyperplane $B_4$\cite{Gasparotto:2024bku}, we have following subset of alphabet 
\begin{align}
\Phi_{(2), \text{bubble}}^{(1)}= &\left\{  X_1-Y_1-Y_2, X_1-Y_1+Y_2,X_1+Y_1+Y_2 \, , \right. \nonumber
\\
& \left. X_2-Y_1-Y_2, X_2-Y_1+Y_2,X_2+Y_1+Y_2 \right. \, ,\nonumber
\\
& \left. X_1+X_2,X_1+X_2+2Y_2 \right\} \, . 
\end{align}
 Exchanging $Y_1$ and $Y_2$, we will get another subset of alphabet $\Phi_{(2), \text{bubble}}^{(2)}$, which is related to the hyperplane $B_5$. We can see that the alphabet of one loop bubble correlators is the union of these two subsets, 
 \begin{align}
 \Phi_{(2), \text{bubble}}=\Phi_{(2), \text{bubble}}^{(1)} \bigcup \Phi_{(2), \text{bubble}}^{(2)} \, .
 \end{align}
 
\begin{figure}[h]
    \centering
    \includegraphics[scale=0.5]{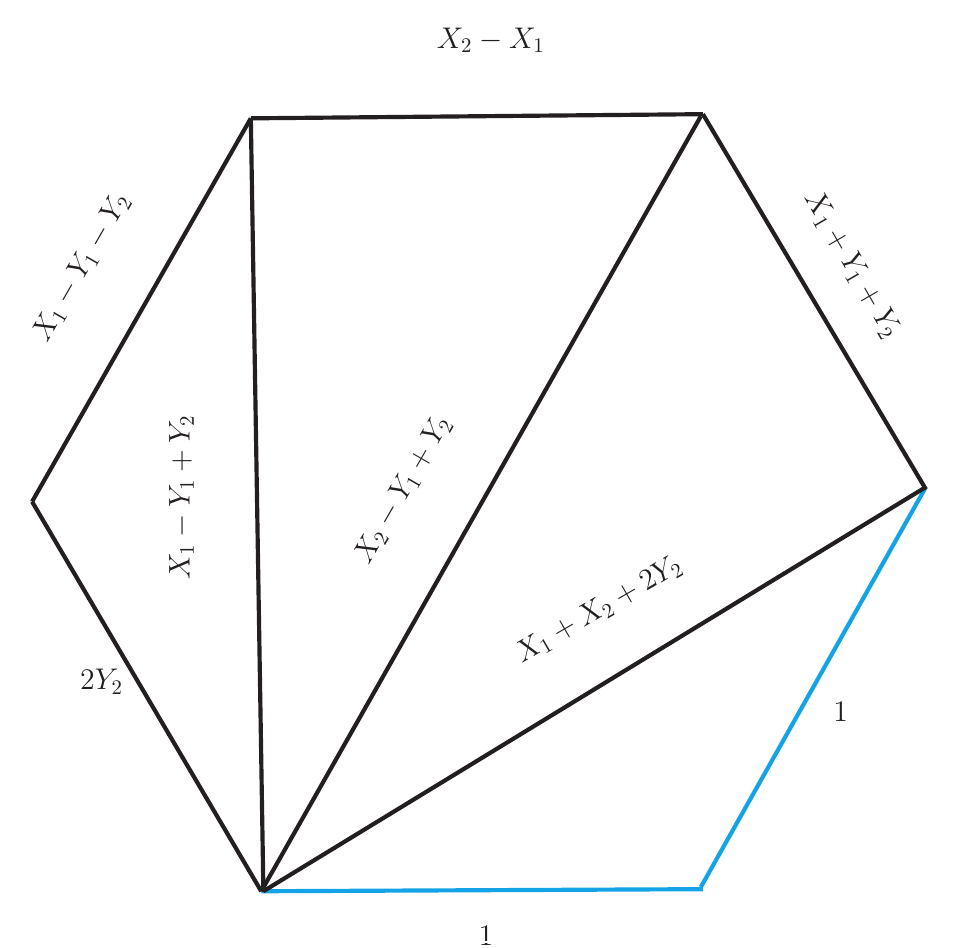}
    \caption{Embedding the letters of $\Phi_{(2), \text{bubble}}^{(1)}$ into a 6-gon.}
    \label{fig:bubble}
\end{figure}

 Let's define following scaleless variables
 \begin{align}
 z_1=\frac{X_1-Y_1-Y_2}{2Y_2} \, , \quad z_2=\frac{X_2-Y_1-Y_2}{2Y_2} \, , \quad z_3=\frac{X_1+X_2}{2Y_2} \, ,
 \end{align}
 the subset $\Phi_{(2), \text{bubble}}^{(1)}$ will be transformed into
 \begin{align}
 \tilde{ \Phi}_{(2), \text{bubble}}^{(1)}=  \left\{z_1,1+z_1,z_2,1+z_2,z_3,1+z_3,z_1-z_3,z_2-z_3\right\} \, .
 \end{align}
 We can identify that $\tilde{ \Phi}_{(2), \text{bubble}}^{(1)}$ is an $A_3$ cluster algebra, except for the missing letter $z_1-z_2$. We find that the letters in $ \Phi_{(2), \text{bubble}}^{(1)}$ can be embedded in a  6-gon, as illustrated in Fig.~\ref{fig:bubble}.  Similar to the tree-level ladder cosmological correlators, the sum edges in 6-gon $2Y_2$, $X_1-Y_1-Y_2$, $(X_2-X_1)$ and $(X_1+Y_1+Y_2)$ 
 satisfy the relation:
\begin{align}
2Y_2 + (X_1 - Y_1 - Y_2) + (X_2 - X_1) + (X_1 + Y_1 + Y_2) = X_1 + X_2 + 2Y_2,
\end{align}
 which corresponds to partition of a line segment of length $(X_1+X_2+2Y_2)$. Notably, the edge $X_2-X_1$ represents a nontrivial new feature arising in the loop integral. 

Since we can obtain the other subset of alphabet by exchanging $Y_1$ and $Y_2$, the subset $\Phi_{(2), \text{bubble}}^{(2)}$ also forms an $A_3$ cluster algebra. This raises an intriguing question: whether these two $A_3$ cluster algebras can be embedded into a larger cluster algebra structure.

\clearpage

\bibliographystyle{JHEP}

\bibliography{bibcf.bib}

\end{document}